%% file: main.tex
\documentclass[conference]{IEEEtran}
\IEEEoverridecommandlockouts

\usepackage{cite}
\usepackage{amsmath,amssymb,amsfonts}
\usepackage{algorithmic}
\usepackage{graphicx}
\usepackage{textcomp}
\usepackage{xcolor}
\usepackage{booktabs}
\usepackage{multirow}

\def\BibTeX{{\rm B\kern-.05em{\sc i\kern-.025em b}\kern-.08em
    T\kern-.1667em\lower.7ex\hbox{E}\kern-.125emX}}

\title{A Survey of Synchronization Technologies for Low-power Backscatter Communication}

\author{
\IEEEauthorblockN{Wenyuan Jiang, Shuo Guo}
\IEEEauthorblockA{Department of Computer Science and Technology, University of Science and Technology of China\\
Email: jiangwenyuan@mail.ustc.edu.cn}
}

\begin{document}

\maketitle

\begin{abstract}
Synchronization is a fundamental enabler for low-power backscatter communication systems, where passive or semi-passive tags modulate ambient RF signals for ultra-low-power data transfer. In this survey, we review recent advances in synchronization techniques across Bluetooth Low Energy (BLE), Long-Term Evolution (LTE), and WiFi-based backscatter platforms. We categorize existing methods by their synchronization granularity, accuracy, compatibility, and power cost. We then compare representative systems including PassiveBLE, Bitalign, LScatter, SyncLTE, LiTEfoot, SyncScatter, and BiScatter, highlighting design trade-offs and performance metrics. Furthermore, we delve into the trade-offs between high throughput and low power synchronization, examining key approaches and challenges such as the balance between throughput, synchronization accuracy, and power consumption in various backscatter systems. Finally, we discuss open challenges and outline future directions toward scalable, secure, and ultra-low-power backscatter synchronization.
\end{abstract}


\section{Introduction}

The growing demand for battery-free and ultra-low-power wireless communication has fueled the development of backscatter systems. These systems enable passive tags to reflect and modulate ambient RF signals emitted by commodity devices, such as WiFi routers, BLE beacons, or LTE base stations. A key challenge in backscatter communication lies in \textit{synchronization}—ensuring that the tag’s reflected signal aligns with the timing of the ambient carrier in a manner that can be reliably decoded by commodity receivers.

Unlike traditional radios, backscatter tags often lack active oscillators, processors, or even continuous power supply, making precise synchronization extremely challenging. Errors in synchronization can lead to inter-symbol interference, frame misalignment, or complete packet loss. Synchronization must thus be performed with high precision, low energy, and often without access to protocol internals.

This paper surveys the current landscape of synchronization technologies designed for backscatter systems. We focus on three major RF ecosystems: BLE, LTE, and WiFi. We analyze several systems—PassiveBLE, Bitalign, LScatter, SyncLTE, LiTEfoot, SyncScatter, and BiScatter—which each tackle synchronization under different constraints and performance targets.

We also explore the trade-off between high throughput and low power synchronization in backscatter communication. Achieving high throughput typically involves more complex signal processing and higher modulation accuracy, which increases power consumption. Conversely, ensuring synchronization accuracy, especially in high-frequency environments, requires precise clock synchronization mechanisms, which also further increase power consumption. These trade-offs are critical in the design of efficient backscatter systems, where minimizing power consumption without sacrificing performance is essential.

We organize the survey as follows: Section II reviews BLE-based techniques. Section III covers LTE-based methods. Section IV discusses WiFi-based designs. Section V presents a cross-technology comparison. Section VI highlights the trade-offs between high throughput and low power synchronization. Section VII discusses open challenges, and Section VIII concludes the paper.
\input{ble}
\input{lte}
\input{wifi}
\input{comparison}
\input{trade-offs}

\input{challenges}
\input{conclusion}

\bibliographystyle{IEEEtran}
\bibliography{reference}

\end{document}

%% file: ble.tex
\section{BLE-based Synchronization Technologies}
Bluetooth Low Energy (BLE) backscatter has gained significant traction due to BLE's widespread deployment in consumer electronics, wearables, and mobile devices. Synchronization in BLE backscatter is particularly challenging due to BLE's frequency hopping spread spectrum (FHSS), packet-level variability, and low-duty-cycle nature. In this section, we discuss two state-of-the-art BLE-based backscatter systems—\textbf{PassiveBLE} and \textbf{Bitalign}—and analyze their design goals, synchronization techniques, and implementation trade-offs.

\subsection{PassiveBLE: Commodity-Compatible BLE Backscatter}
\textbf{PassiveBLE} \cite{passiveble} aims to bridge the gap between BLE backscatter and full compatibility with commodity BLE stacks. Unlike prior systems that were limited to advertising-packet-only communication, PassiveBLE supports full BLE connections using both control and data channels.

\textbf{Key Design Features:}
\begin{itemize}
    \item \textbf{Symbol-level Synchronization via Frequency Difference Detection:} PassiveBLE introduces a novel approach to synchronization based on detecting frequency differences between adjacent BLE symbols. This is motivated by the observation that frequency features are more stable than signal envelope in real-world wireless propagation, leading to reduced jitter and increased alignment precision.
    
    To implement this without an on-board local oscillator (LO), PassiveBLE uses surface acoustic wave (SAW) filters to delay incoming symbols and mix them with real-time signals. The beat frequency reflects the differential frequency, enabling synchronization without high-frequency analog components.

    \item \textbf{Distributed Channel Coding and RF XOR:} BLE packet encoding involves CRC and whitening with dynamically negotiated parameters (e.g., CRC seeds, whitening polynomials). PassiveBLE offloads this burden to the excitor, which pre-encodes the RF waveform. The tag merely performs an XOR operation over this waveform to generate valid BLE-compliant packets.

    \item \textbf{BLE Connection Scheduler:} BLE connections require maintaining state (e.g., connection intervals, hopping patterns, access addresses). PassiveBLE repurposes a modified BLE transceiver as the excitor to manage all connection-level control, freeing the tag from control-plane overhead.
\end{itemize}

\textbf{Performance Highlights:}
\begin{itemize}
    \item Over 99.9\% success rate in BLE connection establishment across varying channel conditions.
    \item Throughput of 532 kbps in LE 1M mode and 974 kbps in LE 2M mode, a 63.3$\times$ improvement over traditional BLE backscatter.
    \item Power consumption of 491 $\mu$W (prototype) and 9.9 $\mu$W (ASIC), demonstrating real-world feasibility.
\end{itemize}

\textbf{Insights:} PassiveBLE highlights the feasibility of achieving full BLE protocol compatibility without violating the BLE specification. Its modular offloading strategy—synchronization, encoding, and control—is essential for scalable BLE tag design. However, this comes at the cost of more complex excitor logic and greater reliance on customized hardware support.

\subsection{Bitalign: Bit-Accurate Synchronization for BLE Backscatter}
\textbf{Bitalign} \cite{bitalign} focuses on accurate symbol-level synchronization in ambient BLE environments, especially when the BLE source is uncontrollable and exhibits inconsistencies in preamble structure. Modern BLE chipsets such as the Nordic 52833 introduce pilot tones and edge jitter, causing legacy backscatter assumptions to fail.

\textbf{Key Contributions:}
\begin{itemize}
    \item \textbf{Identification-based Synchronization:} Bitalign first constructs a fingerprint model of the BLE signal's envelope, capturing key inflection points and energy transitions. By matching this fingerprint in runtime against live signals, the tag can infer the true start of packet boundaries with high probability.
    
    \item \textbf{Matching-based Refinement:} After coarse synchronization, a finer-grained alignment is achieved by signal template matching at the bit level. This step compensates for preamble distortions introduced by chip-specific behavior.

    \item \textbf{BLE Header Prediction:} Bitalign proposes an intelligent mechanism for reconstructing BLE packet headers, allowing correct whitening and CRC application. This enables the tag to generate modulated packets that match BLE protocol requirements without full decoding.
\end{itemize}

\textbf{Performance Highlights:}
\begin{itemize}
    \item Synchronization error reduced from 58.5\% to 5\% (identification-based) and further to 0.5\% with matching-based alignment.
    \item Achieves up to 1.98 Mbps throughput with commodity BLE excitations.
    \item Compatible with BLE 5.0 PHY modes and validated across multiple chipsets.
\end{itemize}

\textbf{Comparison to PassiveBLE:} While PassiveBLE provides full protocol compliance via custom hardware and scheduling, Bitalign excels in opportunistic settings where the BLE source cannot be controlled. It adapts to real-world BLE signal variations and enables high-throughput backscatter without requiring tight excitor-tag coupling.

\textbf{Open Questions:} BLE-based backscatter remains a fast-evolving domain. Outstanding challenges include tag scalability under BLE's frequency-hopping scheme, coexistence with dense BLE networks, and BLE 5.2+ support (e.g., Isochronous Channels). Hybrid systems that combine PassiveBLE's structure with Bitalign's adaptability could represent the next frontier.

%% file: lte.tex
\section{LTE-based Synchronization Technologies}
Long-Term Evolution (LTE) signals are an attractive ambient carrier source for backscatter communication due to their wide coverage, strong signal structure, and stable periodic synchronization signals. However, exploiting LTE for backscatter synchronization involves overcoming significant challenges, such as complex frame structures, bursty subframe activity, and the high-frequency overhead of downlink control. In this section, we discuss three representative systems—\textbf{LScatter}, \textbf{SyncLTE}, and \textbf{LiTEfoot}—each addressing synchronization from different design perspectives.

\subsection{LScatter: Exploiting LTE Synchronization Signals for Throughput-Oriented Backscatter}
LScatter \cite{lscatter} represents one of the earliest efforts to perform high-throughput backscatter using ambient LTE traffic. Unlike BLE or WiFi systems, LTE frames contain built-in synchronization channels such as the Primary Synchronization Signal (PSS) and Secondary Synchronization Signal (SSS), which repeat every 5 ms and serve as stable anchors for passive receivers.

\textbf{Core Contributions:}
\begin{itemize}
    \item \textbf{Symbol-level Synchronization via PSS Detection:} LScatter uses energy and correlation-based detection of LTE PSS sequences to establish timing reference points. This enables the tag to synchronize at sub-symbol resolution.
    \item \textbf{Phase Offset Correction:} To mitigate the effects of fading and oscillator drift, LScatter leverages LTE reference signals (RS) to estimate and correct residual phase errors in the backscattered waveform.
    \item \textbf{Subcarrier-Aligned Modulation:} By aligning modulation to OFDM subcarriers, LScatter ensures that backscattered packets remain decodable by commodity LTE receivers in out-of-band (OOB) monitoring mode.
\end{itemize}

\textbf{Performance:}
\begin{itemize}
    \item Achieves up to 13.63 Mbps throughput by leveraging dense LTE PDSCH regions.
    \item Supports link distances of 8--15 meters in indoor environments.
    \item However, does not ensure compatibility with legacy LTE receivers due to non-compliant frame insertion.
\end{itemize}

\textbf{Limitations:} LScatter's synchronization is susceptible to signal burst variability and cannot guarantee decoding on unmodified UEs. Its high-throughput focus trades off compatibility.

\subsection{SyncLTE: Precision Synchronization with Standard Compatibility}
SyncLTE \cite{synclte} aims to provide robust synchronization with full LTE protocol compliance. Unlike LScatter, SyncLTE explicitly avoids modifying LTE control channels and modulates only data symbols in the PDSCH.

\textbf{Key Techniques:}
\begin{itemize}
    \item \textbf{Template-based Correlation:} SyncLTE constructs correlation templates from publicly known LTE signal structures (PSS, SSS, RS) and uses a sliding window to detect symbol boundaries with high fidelity.
    \item \textbf{Symbol-Retentive Backscatter:} It employs single-symbol amplitude modulation schemes (e.g., OOK or ASK) to embed data without violating LTE field integrity.
    \item \textbf{Commodity Compatibility:} SyncLTE is explicitly evaluated with off-the-shelf smartphones and LTE modems in monitor mode, demonstrating compatibility with legacy LTE receivers.
\end{itemize}

\textbf{Results:}
\begin{itemize}
    \item Synchronization error reduced by 22.4$\times$ vs. LScatter (80\textsuperscript{th} percentile).
    \item Supports uplink ranges of up to 36 meters.
    \item Throughput up to 400 bps (QPSK) and 200 bps (BPSK) under standard LTE PHY constraints.
\end{itemize}

\textbf{Strengths:} SyncLTE introduces precise synchronization and standard-aware modulation, making it ideal for medical or industrial deployments requiring legacy LTE integration.

\subsection{LiTEfoot: Passive Synchronization for Localization and Communication}
LiTEfoot \cite{litefoot} takes a distinct approach—using synchronization primarily for localization, but with principles extensible to backscatter communication. Its novelty lies in performing synchronization without downconverting signals or using expensive ADCs.

\textbf{Architectural Innovations:}
\begin{itemize}
    \item \textbf{Nonlinear Spectrum Folding:} By applying envelope detection to the wideband LTE spectrum, LiTEfoot folds subcarrier components (e.g., PSS, SSS) into baseband via intermodulation.
    \item \textbf{Wideband Scanning in 10 ms:} This enables real-time detection of cell identities across a 3 GHz LTE spectrum window using only 0.9 mJ.
    \item \textbf{Envelope-domain Correlation:} Rather than decoding packets, it performs correlation on envelope shapes to detect synchronization sequences.
\end{itemize}

\textbf{Achievements:}
\begin{itemize}
    \item 22 m median localization error in urban settings.
    \item Energy consumption of 40 $\mu$J (simulated CMOS) or 0.8 mJ (PCB prototype).
    \item Design extensible to 5G NR and NB-IoT synchronization.
\end{itemize}

\textbf{Interpretation:} While not strictly a communication system, LiTEfoot's architecture shows how to build ultra-low-power front-ends capable of high-rate spectrum monitoring and passive synchronization.

\subsection{Discussion and Synthesis}
\begin{itemize}
    \item \textbf{LScatter} prioritizes data rate and physical-layer performance, sacrificing compliance.
    \item \textbf{SyncLTE} enforces LTE standard compatibility with lower data rates but better reliability and coexistence.
    \item \textbf{LiTEfoot} enables passive spectrum sensing and synchronization without RF downconversion, offering an attractive model for battery-free tags.
\end{itemize}

Together, these works illustrate a spectrum of LTE-based backscatter designs, with trade-offs across synchronization accuracy, protocol compatibility, power consumption, and real-world deployability.

%% file: wifi.tex
\section{WiFi-based Synchronization Technologies}
WiFi backscatter communication is particularly attractive for indoor environments due to the ubiquity of 802.11 infrastructure and the availability of continuous high-frequency transmissions. However, achieving precise synchronization in WiFi-based backscatter is challenging because of the bursty and unpredictable traffic patterns, variable preamble structures, and complex PHY-layer modulation. This section reviews two influential systems—\textbf{SyncScatter} and \textbf{HitchHike}—and contrasts their synchronization mechanisms, capabilities, and limitations.

\subsection{SyncScatter: Symbol-Level Synchronization for Commodity WiFi}
SyncScatter \cite{syncscatter} represents a significant advancement in WiFi backscatter by enabling symbol-aligned synchronization with commercial 802.11b/g/n access points. Prior systems relied solely on energy detection, which suffers from microsecond-scale jitter. SyncScatter instead introduces a hierarchical synchronization architecture combining low-power sensing with fine-grained RF processing.

\textbf{Key Technical Components:}
\begin{itemize}
    \item \textbf{Two-Stage Synchronization Frontend:} The first stage is a passive wake-up detector that monitors channel activity and triggers a second-stage high-precision amplifier and digitizer only upon detection. This architecture balances energy efficiency and synchronization accuracy.

    \item \textbf{Symbol-Timing Alignment:} SyncScatter utilizes preamble detection and RF envelope tracking to align backscatter modulation with the exact symbol boundaries of OFDM or DSSS WiFi transmissions. This mitigates inter-symbol interference (ISI) and enhances decodability.

    \item \textbf{Hardware Co-design with ASIC:} A custom chip integrates all RF synchronization logic, enabling 150 ns timing accuracy while consuming only 30 $\mu$W.
\end{itemize}

\textbf{Performance:}
\begin{itemize}
    \item Achieves reliable backscatter communication at ranges up to 30 meters.
    \item Supports 500 kbps data rate using 802.11b PHY.
    \item Maintains packet error rate below 1\% under normal indoor interference.
\end{itemize}

\textbf{Strengths:} SyncScatter provides a scalable solution for multi-tag environments and is well-suited to energy-harvesting use cases such as smart homes or wearables.

\subsection{HitchHike: Opportunistic Synchronization via Energy Detection}
HitchHike \cite{hitchhike} is an earlier design that allows backscatter transmission using commodity WiFi preambles. It requires no synchronization signal from the excitor and instead uses the rising edge of WiFi energy to trigger backscatter.

\textbf{Mechanism Overview:}
\begin{itemize}
    \item \textbf{Energy Thresholding:} Tags detect the energy burst of an incoming packet and initiate modulation at a fixed offset.
    \item \textbf{Asynchronous Modulation:} Without access to symbol or frame alignment, tags modulate continuously during detected bursts.
    \item \textbf{Receiver Customization:} Decoding is performed by modified receivers capable of interpreting the backscatter waveform amidst the WiFi signal.
\end{itemize}

\textbf{Limitations:}
\begin{itemize}
    \item Synchronization jitter around 2 $\mu$s limits achievable throughput and increases BER.
    \item Sensitive to channel fading and mobility.
    \item Requires proximity to both transmitter and receiver (typically under 6 meters).
\end{itemize}

\subsection{Comparative Analysis and Insights}
\begin{itemize}
    \item \textbf{Precision:} SyncScatter achieves sub-microsecond accuracy, enabling reliable OFDM/DSSS synchronization. HitchHike operates on millisecond granularity, suitable only for coarse packet timing.
    \item \textbf{Energy:} Both systems target low power, but SyncScatter's two-stage architecture allows power gating, enabling sub-100 $\mu$W designs.
    \item \textbf{Compatibility:} SyncScatter is compatible with unmodified APs and receivers. HitchHike requires receiver-side customization, reducing deployability.
\end{itemize}

\textbf{Open Challenges:}
\begin{itemize}
    \item Supporting high-throughput backscatter under 802.11ac/ax with very short guard intervals.
    \item Enabling multi-user synchronization when tags share WiFi bursts.
    \item Reducing latency in symbol detection under dynamic channel loads.
\end{itemize}

Overall, SyncScatter exemplifies how architectural co-design and hierarchical synchronization unlock practical WiFi backscatter. Future work may explore deep learning-based synchronization or cross-technology fusion (e.g., BLE-WiFi) to further enhance robustness.

%% file: comparison.tex
\section{Cross-Technology Comparison and Unified Insights}
With BLE, LTE, and WiFi each offering ambient sources for backscatter communication, understanding their synchronization trade-offs is crucial for selecting or designing systems based on application needs. In this section, we offer a detailed cross-layer analysis of representative systems, focusing on synchronization performance, power cost, throughput, and deployment flexibility.

\subsection{Feature Comparison Matrix}
Table~\ref{tab:compare} highlights core differences across BLE, LTE, and WiFi-based backscatter synchronization systems.

\begin{table*}[htbp]
\caption{Comparison of Synchronization Technologies in Backscatter Systems}
\label{tab:compare}
\centering
\begin{tabular}{|c|c|c|c|c|c|c|}
\hline
\textbf{System} & \textbf{Protocol} & \textbf{Sync Method} & \textbf{Precision} & \textbf{Throughput} & \textbf{Compatibility} & \textbf{Power} \\
\hline
PassiveBLE~\cite{passiveble} & BLE & Freq-diff detection & 1 $\mu$s & 974 kbps & Full BLE & 9.9 $\mu$W (ASIC) \\
Bitalign~\cite{bitalign} & BLE & Feature Matching & 0.5\% BER & 1.98 Mbps & Partial & 10s of $\mu$W \\
SyncLTE~\cite{synclte} & LTE & Template correlation & 15 $\mu$s & 200--400 bps & Legacy UEs & 100 $\mu$W \\
LScatter~\cite{lscatter} & LTE & PSS + RS tracking & $\sim$50 $\mu$s & 13.63 Mbps & Partial & Moderate \\
LiTEfoot~\cite{litefoot} & LTE & Spectrum folding & ms-level (envelope) & N/A (localization) & N/A & 0.8 mJ \\
SyncScatter~\cite{syncscatter} & WiFi & Envelope + symbol tracking & 150 ns & 500 kbps & 802.11 APs & 30 $\mu$W \\
HitchHike~\cite{hitchhike} & WiFi & Energy detection & 2 $\mu$s & $<$11 kbps & Modified Rx & 10s of $\mu$W \\
\hline
\end{tabular}
\end{table*}

\subsection{Synchronization Trade-offs}
\textbf{Accuracy vs. Energy:} Systems like SyncScatter and PassiveBLE achieve sub-microsecond precision but require careful analog frontend design and ASIC support. In contrast, envelope-based methods like LiTEfoot prioritize passive listening and energy minimization.

\textbf{Throughput vs. Compliance:} LScatter attains high throughput by modulating LTE symbols directly but violates LTE protocol semantics. SyncLTE and PassiveBLE maintain compliance at the cost of reduced bit rates. This reveals a tension between physical-layer efficiency and standard compatibility.

\textbf{Complexity vs. Robustness:} Systems with tighter timing requirements (e.g., Bitalign) must compensate for oscillator jitter and vendor-specific PHY variations, adding signal processing complexity. Simpler systems (e.g., HitchHike) offer easier deployment but at the cost of range, BER, and precision.

\subsection{Protocol-Level Observations}
\begin{itemize}
    \item \textbf{BLE:} Offers controlled connection intervals and well-defined packet structure but suffers from rapid frequency hopping.
    \item \textbf{LTE:} Provides rich synchronization references (PSS, SSS, RS) and time predictability, but integration with commercial UEs is difficult.
    \item \textbf{WiFi:} Benefits from frequent bursts and dense deployments, but short preambles and variable inter-frame gaps complicate synchronization.
\end{itemize}

\subsection{Application-Level Recommendations}
\begin{itemize}
    \item \textbf{Energy-constrained localization:} LiTEfoot is ideal for GPS-free tracking under tight energy budgets.
    \item \textbf{BLE-compatible health monitoring:} PassiveBLE provides a reliable bridge between ultra-low-power sensing and BLE smartphones.
    \item \textbf{Smart home connectivity:} SyncScatter offers symbol-aligned transmission with minimal hardware and wide WiFi coverage.
    \item \textbf{High-rate backscatter for video/sensors:} LScatter or Bitalign may be suitable, with excitor modifications and strong processing.
\end{itemize}

\subsection{Toward Unified Frameworks}
Future research may explore hybrid systems that dynamically switch synchronization references (e.g., BLE + LTE), leverage machine learning to predict packet boundaries under noisy conditions, or coordinate multiple tags via cooperative timing protocols. The long-term goal is a unified synchronization stack supporting diverse protocols, timing granularities, and energy constraints across billions of edge devices.

%% file: trade-offs.tex
\section{Trade-off Between High Throughput and Low Power Synchronization}

\subsection{Balancing Throughput, Accuracy, and Power Consumption}

In low-power backscatter communication systems, finding an optimal balance between throughput, synchronization accuracy, and power consumption is a key design challenge. Achieving high throughput often requires more complex signal processing and higher modulation accuracy, which increases power consumption. On the other hand, ensuring synchronization accuracy, especially in high-frequency environments, requires precise clock synchronization mechanisms, further increasing power consumption.

\subsubsection{Key Approaches and Trade-offs}

Several systems have been proposed to address the trade-off between these factors:

\begin{itemize}
    \item \textbf{PilotScatter}\cite{pilotscatter}: Modulates the phase and amplitude of pilot symbols for data transmission and employs differential demodulation methods to reduce reliance on high-precision clocks. This design achieves high throughput while maintaining reasonable power consumption and supports 16-QAM modulation.
    \item \textbf{SymbolBack}\cite{symbolback}: Introduces a symbol-level backscatter communication method by precisely controlling the media access control (MAC) payload, improving synchronization accuracy through reference symbols. However, as throughput increases, the trade-off between accuracy and throughput leads to higher power consumption.
    \item \textbf{EchScatter}\cite{echscatter}: Uses ZigBee signals as excitation sources and applies phase modulation for symbol-level modulation, significantly improving throughput. It optimizes synchronization accuracy under low power using an average-energy detection synchronization method.
\end{itemize}

These systems demonstrate that while throughput and synchronization accuracy are often in conflict, trade-offs can be optimized, particularly in low-power environments.

\subsubsection{Challenges in Millimeter-Wave Backscatter}

Millimeter-wave backscatter communication systems take advantage of the abundant bandwidth resources in the millimeter-wave frequency band. However, they face significant challenges due to signal attenuation. \cite{chen2024survey} provides a comprehensive overview of these systems, highlighting the power consumption challenges and suggesting intelligent beamforming techniques to reduce energy use.

\subsubsection{Low-Power Synchronization for Healthcare Applications}

\textbf{Low-Power Synchronization for Multi-IMU WSNs}\cite{cappelle} introduces a multi-IMU wireless synchronization platform designed for low-power healthcare applications. The system achieves sub-microsecond synchronization accuracy while maintaining energy consumption as low as 74.8 J/h, showing how precision and low power can be balanced effectively for wearable health devices.

\subsubsection{Feature Management in Software Product Lines}

\cite{michelon2023analysis} discusses the management of feature revisions in preprocessor-based software product lines, emphasizing the importance of balancing performance and resource consumption. The principles discussed are applicable to hardware and software co-design, especially in low-power backscatter communication systems, ensuring efficient synchronization without compromising system performance.

\subsubsection{Radar-Based Backscatter Communication}

Finally, \textbf{BiScatter}\cite{okubo2024integrated} integrates radar backscatter with communication functionality, proposing an innovative two-way communication and sensing system. This system resolves the traditional trade-off between synchronization accuracy and throughput, enabling smooth data transmission and precise location tracking.

\subsection{Impact of Application Scenarios}

The trade-off between high throughput and low power synchronization varies depending on the specific application scenario. Below are some key examples:

\subsubsection{Medical and Smart Home Applications}

In medical and smart home applications, high reliability and precision are essential within limited power budgets. Throughput is less critical compared to synchronization accuracy and power management. For instance, SymbolBack is ideal for applications such as intelligent health monitoring, where high synchronization accuracy is necessary but throughput demands are lower.

\subsubsection{Internet of Things (IoT) Applications}

In IoT applications, where many devices are involved, flexible decisions need to be made between throughput and synchronization accuracy:

\begin{itemize}
    \item \textbf{DBscatter}\cite{wang2024enabling}: Supports dual-band Wi-Fi backscatter communication (2.4 GHz and 5 GHz), offering higher throughput while reducing power consumption. The dual-band design improves signal reliability and reduces interference, making it ideal for IoT applications, especially in smart homes.
\end{itemize}

This flexibility allows IoT systems to balance power consumption and throughput based on environmental needs.

\subsubsection{High Bandwidth Applications}

For applications requiring high bandwidth, such as video streaming or augmented reality (AR), high throughput becomes a critical design requirement. BiScatter integrates radar backscatter with communication functionality, providing both data transmission and precise location services. The system ensures smooth data transmission while also enabling location tracking in scenarios that require both.

%% file: challenges.tex
\section{Open Challenges and Future Directions}
While significant progress has been made in developing synchronization techniques for backscatter communication, several open challenges remain that hinder large-scale, robust, and truly battery-free deployment. These challenges span physical-layer limitations, system-level coordination, and integration with emerging wireless standards.

\subsection{Challenge 1: Multi-Tag Synchronization and Network Scalability}
Current systems are typically optimized for point-to-point links or small-scale testbeds. However, practical deployments often involve dozens to hundreds of tags operating in the same RF space.
\begin{itemize}
    \item \textbf{Issue:} Coordinating timing among multiple passive tags without creating destructive interference or missed synchronization opportunities.
    \item \textbf{Direction:} Future work may explore distributed synchronization protocols using low-power contention mechanisms, token-based wake-up schemes, or RF-coded scheduling.
\end{itemize}

\subsection{Challenge 2: Adaptive Synchronization Under Dynamic Interference}
WiFi, BLE, and LTE operate in increasingly congested environments. Synchronization signals may suffer from multi-path fading, burst collisions, or temporal drift.
\begin{itemize}
    \item \textbf{Issue:} Static correlation thresholds and fixed timing models fail in fluctuating SNR and traffic conditions.
    \item \textbf{Direction:} Intelligent signal classifiers or lightweight machine learning models could provide adaptive synchronization, selectively retraining templates or adjusting detection thresholds.
\end{itemize}

\subsection{Challenge 3: Ultra-Low-Power Hardware Realization}
While envelope detectors and SAW filters reduce circuit complexity, full synchronization circuits (e.g., ADCs, PLLs) remain energy-expensive.
\begin{itemize}
    \item \textbf{Issue:} Most current designs still depend on wake-up amplifiers or digital control blocks.
    \item \textbf{Direction:} Explore novel analog-only circuits, asynchronous logic, and energy-aware design patterns to bring synchronization into the nanojoule regime.
\end{itemize}

\subsection{Challenge 4: Cross-Technology and Multi-Band Synchronization}
Each protocol offers benefits—BLE for simplicity, LTE for periodicity, WiFi for density—but combining them is non-trivial.
\begin{itemize}
    \item \textbf{Issue:} Multi-protocol synchronization requires reconciling incompatible timing models, symbol rates, and frequency plans.
    \item \textbf{Direction:} Cross-band synchronization strategies could leverage shared reference events (e.g., cellular base station beacons) or unified timing anchors in multi-radio platforms.
\end{itemize}

\subsection{Challenge 5: Secure Synchronization Against Spoofing and Evasion}
Backscatter tags lack the computational resources to validate signal authenticity.
\begin{itemize}
    \item \textbf{Issue:} Malicious actors can transmit fake synchronization pulses, causing tags to misalign or jam legitimate communication.
    \item \textbf{Direction:} Lightweight cryptographic hashes, physical-layer watermarking, or redundancy-based validation could be explored.
\end{itemize}

\subsection{Challenge 6: Benchmarking and Reproducibility}
Despite rapid progress, there is no standard framework for evaluating backscatter synchronization systems.
\begin{itemize}
    \item \textbf{Issue:} Papers report different metrics (e.g., BER, timing jitter, detection rate) using ad hoc hardware setups.
    \item \textbf{Direction:} Establish shared open-source testbeds, reference signal datasets, and standard synchronization error definitions (e.g., P50, P90 latency).
\end{itemize}

\subsection{Looking Forward}
To unlock ubiquitous and seamless backscatter deployment, future systems must unify low-power hardware, adaptive synchronization logic, and secure multi-tag protocols. Bridging cross-layer boundaries—combining RF design, MAC scheduling, and learning-based adaptation—will be key to enabling scalable, reliable, and intelligent synchronization in the next generation of ambient IoT systems.

%% file: conclusion.tex
\section{Conclusion}
Synchronization is a cornerstone for enabling reliable and efficient communication in low-power backscatter systems. This survey has reviewed state-of-the-art synchronization technologies across BLE, LTE, and WiFi-based backscatter platforms. From frequency-difference detection in PassiveBLE to symbol-aligned template correlation in SyncLTE, and hierarchical wake-up protocols in SyncScatter, these systems reveal a spectrum of trade-offs between precision, energy efficiency, and compatibility.

We presented a cross-technology taxonomy of synchronization methods, compared core design principles, and identified key trade-offs and architectural insights. Despite the diversity of hardware and protocols, unifying themes emerge around decoupling control logic, leveraging ambient periodicity, and co-designing low-power synchronization circuits.we discussed various systems such as PilotScatter, SymbolBack, and EchScatter, which demonstrate how these trade-offs can be addressed through innovative techniques that balance synchronization precision with power efficiency.

Looking ahead, addressing open challenges—such as scalability, adaptive robustness, and standard unification—will be essential to unlock the full potential of ubiquitous, long-range, and battery-free communication. With continued progress in RF design, machine learning, and cross-layer integration, synchronization may become a solved problem in future backscatter-enabled IoT ecosystems.

%% file: main.bbl
\begin{thebibliography}{10}
\providecommand{\url}[1]{#1}
\csname url@samestyle\endcsname
\providecommand{\newblock}{\relax}
\providecommand{\bibinfo}[2]{#2}
\providecommand{\BIBentrySTDinterwordspacing}{\spaceskip=0pt\relax}
\providecommand{\BIBentryALTinterwordstretchfactor}{4}
\providecommand{\BIBentryALTinterwordspacing}{\spaceskip=\fontdimen2\font plus
\BIBentryALTinterwordstretchfactor\fontdimen3\font minus \fontdimen4\font\relax}
\providecommand{\BIBforeignlanguage}[2]{{%
\expandafter\ifx\csname l@#1\endcsname\relax
\typeout{** WARNING: IEEEtran.bst: No hyphenation pattern has been}%
\typeout{** loaded for the language `#1'. Using the pattern for}%
\typeout{** the default language instead.}%
\else
\language=\csname l@#1\endcsname
\fi
#2}}
\providecommand{\BIBdecl}{\relax}
\BIBdecl

\bibitem{passiveble}
H.~Dong, Y.~Wu, F.~Li, W.~Kuang, Y.~He, Q.~Zhang, and W.~Wang, ``Passiveble: Towards fully commodity-compatible ble backscatter,'' in \emph{Proc. ACM MobiCom}, 2025.

\bibitem{bitalign}
Z.~Huang, Z.~Qian, H.~Ma, D.~Pei, and K.~Wu, ``Bitalign: Bit alignment for bluetooth backscatter communication,'' \emph{IEEE Transactions on Mobile Computing}, vol.~23, no.~10, pp. 10\,191--10\,205, 2024.

\bibitem{lscatter}
Z.~Chi, X.~Liu, W.~Wang, Y.~Yao, and T.~Zhu, ``Leveraging ambient lte traffic for ubiquitous passive communication,'' in \emph{Proc. ACM SIGCOMM}, 2020.

\bibitem{synclte}
Y.~Feng, S.~Chen, W.~Xi, S.~Wang, J.~Zhao, and W.~Gong, ``Heartbeating with lte networks for ambient backscatter,'' \emph{IEEE Transactions on Mobile Computing}, vol.~23, no.~5, pp. 4246--4260, 2024.

\bibitem{litefoot}
N.~Garg, A.~Ghosh, and N.~Roy, ``Litefoot: Ultra-low-power localization using ambient cellular signals,'' in \emph{Proc. ACM SenSys}, 2024.

\bibitem{syncscatter}
M.~Dunna, M.~Meng, P.-H. Wang, C.~Zhang, P.~Mercier, and D.~Bharadia, ``Syncscatter: Enabling wifi-like synchronization and range for wifi backscatter communication,'' in \emph{Proc. USENIX NSDI}, 2021.

\bibitem{hitchhike}
V.~Iyer, V.~Talla, B.~Kellogg, S.~Gollakota, and J.~R. Smith, ``Hitchhike: Practical backscatter using commodity wifi,'' in \emph{Proc. ACM SenSys}, 2016.

\bibitem{pilotscatter}
Q.~Wang, J.~Zhao, and W.~Gong, ``Pilotscatter: High-throughput ofdm backscatter via pilot tones,'' \emph{IEEE Transactions on Wireless Communications}, 2024.

\bibitem{symbolback}
S.~Zhang, Y.~Yan, and S.~Wang, ``Symbolback: Symbol-level wifi backscatter communication,'' in \emph{2024 10th International Conference on Big Data Computing and Communications (BigCom)}.\hskip 1em plus 0.5em minus 0.4em\relax IEEE, 2024, pp. 204--210.

\bibitem{echscatter}
J.~Li, S.~Wang, Z.~Xu, W.~Xi, S.~Wang, and W.~Gong, ``Echscatter: Enriching codeword translation for high-throughput ambient zigbee backscatter,'' \emph{ACM Transactions on Sensor Networks}, 2025.

\bibitem{chen2024survey}
W.~Chen, W.~Yang, and W.~Gong, ``A survey of millimeter wave backscatter communication systems,'' \emph{Computer Networks}, vol. 242, p. 110235, 2024.

\bibitem{cappelle}
J.~Cappelle, S.~Goossens, L.~De~Strycker, and L.~Van~der Perre, ``Low-power synchronization for multi-imu wsns,'' \emph{IEEE Embedded Systems Letters}, vol.~16, no.~2, pp. 210--213, 2023.

\bibitem{michelon2023analysis}
G.~K. Michelon, W.~K. Assun{\c{c}}{\~a}o, P.~Gr{\"u}nbacher, and A.~Egyed, ``Analysis and propagation of feature revisions in preprocessor-based software product lines,'' in \emph{2023 IEEE International Conference on Software Analysis, Evolution and Reengineering (SANER)}.\hskip 1em plus 0.5em minus 0.4em\relax IEEE, 2023, pp. 284--295.

\bibitem{okubo2024integrated}
R.~Okubo, L.~Jacobs, J.~Wang, S.~Bowers, and E.~Soltanaghai, ``Integrated two-way radar backscatter communication and sensing with low-power iot tags,'' in \emph{Proceedings of the ACM SIGCOMM 2024 Conference}, 2024, pp. 327--339.

\bibitem{wang2024enabling}
B.~Wang, F.~Zhu, L.~Zhong, M.~Jin, X.~Wang, C.~Chen, X.~Guan, C.~Zhou, and X.~Tian, ``Enabling dual-band wi-fi backscatter,'' \emph{IEEE Transactions on Mobile Computing}, 2024.

\end{thebibliography}
